\documentclass[11pt]{article}
\usepackage{fullpage}
\usepackage[top=0.9in,bottom=1.1in,left=1.1in,right=1.1in]{geometry}
\usepackage{graphicx}
\usepackage{amsmath,amssymb,graphicx,paralist,amsthm}
\RequirePackage{hyperref}
\usepackage{times}
\usepackage{anyfontsize}
\usepackage{subcaption}
\usepackage{authblk}

\usepackage{xcolor}
\usepackage{bbm,bm}
\usepackage{natbib} 
\usepackage{multirow}
\usepackage[ruled]{algorithm2e}
\usepackage{algpseudocode}
\usepackage{float}

\newtheorem{theorem}{Theorem}[section]

\usepackage{xr}
\newtheorem{assumption}{Assumption}
\newcommand{\ind}{\perp\!\!\!\!\perp} %\externaldocument[]{Supp-Transfer-GLM-unblind}
\usepackage{float}
 
\usepackage{graphicx}
\newcommand{\ignore}[1]{}
\title{Mediation analysis with densities as mediators with an application to iCOMPARE trial}
\author{Jingru Zhang, Mathias Basner,
    Christopher W.\ Jones,
    David F. Dinges,\\
    \ Haochang Shou \ and \ Hongzhe Li
\footnote{Jingru Zhang is a postdoctoral fellow in Department of Biostatistics, Epidemiology and Informatics, 
    Mathias Basner is Professor of Psychiatry, 
    Christopher W.\ Jones is a postdoctoral fellow, and 
    David F. Dinges is Professor of Psychiatry in Department of Psychiatry, 
    Haochang Shou is an Assistant Professor, Hongzhe Li is Perelman Professor of Biostatistics (E-mail: \emph{hongzhe@upenn.edu}),  Department of Biostatistics, Epidemiology and Informatics, Perelman School of Medicine, University of Pennsylvania, Philadelphia, PA 19104.}} 

\date{}
\begin{document}

  \maketitle

\begin{abstract}
Physical activity has long been shown to be  associated with biological and physiological performance and risk of diseases. It is of great interest to assess whether the effect of an exposure or intervention on an outcome is mediated through physical activity measured by modern wearable devices such as actigraphy. 
However, existing methods for  mediation analysis focus almost exclusively on mediation variable  that is in the Euclidean space, which cannot be applied directly to the actigraphy data of  physical activity. Such data is best summarized in the form of an histogram or  density. 
In this paper, we extend the structural equation models (SEMs) to the settings where a density  is treated as  the mediator to study the indirect mediation effect of physical activity on an outcome. We provide sufficient conditions for identifying the average causal effects of density 
mediator and present methods for estimating the direct and mediating effects of density on an outcome. 
We apply our method to the data set from the iCOMPARE trial that  compares flexible duty-hour policies  and standard duty-hour policies on interns' sleep related outcomes  to  explore the mediation effect of physical activity on the causal path between  flexible duty-hour policies and sleep related outcomes.

\end{abstract}

%%%%%%%%%%%%%%%%%%%%%%%%%%%%%%%%%%%%%%%%
\section{Introduction and iCOMPARE Trial}
There has been an increased interest in exploring whether a policy would have an effect on psychological and physiological outcomes and in understanding the mechanism that leads to such effects. 
Physical activity has been shown to be associated with biological and physiological performance \citep{activity,activity1}.  One possible link between a policy and biological and physiological performance is through the change of physical activities. A question of  interest is whether a policy influences psychological/physiological variables through physical activity as a mediator.  
This question can be addressed via  mediation analysis, which  provides a mean of  studying the underlying mechanism or process by which a treatment or policy  has an effect on an outcome through a mediator.

\begin{figure}[!ht]
    \centering
    \begin{tabular}{cc}
        \includegraphics[width=0.48\textwidth]{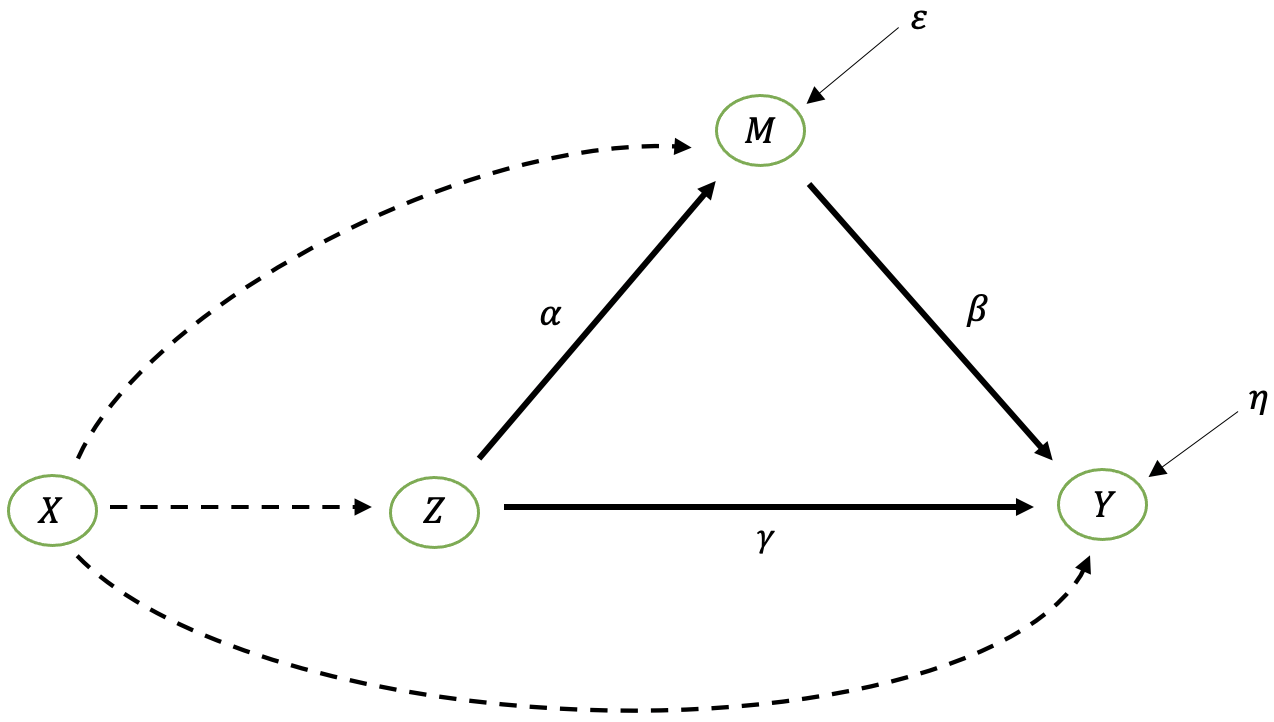}&
        \includegraphics[width=0.48\textwidth]{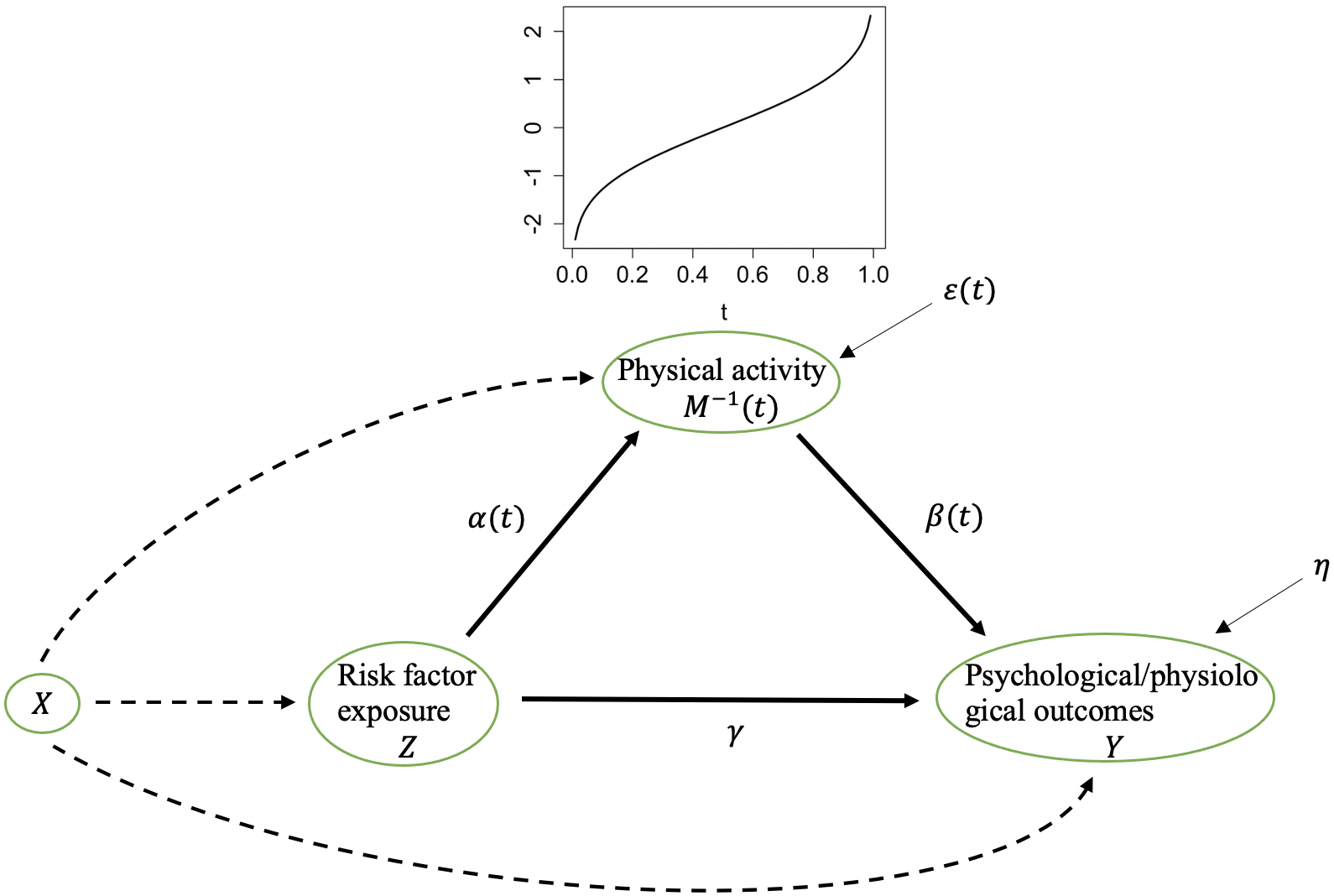}\\
        (a) standard mediation framework&
        (b) density  as a mediator
    \end{tabular}
    \caption{(a): The path diagram representation of the standard mediation framework. The variables $Z$, $Y$, $M$, and the path coefficients $\alpha$, $\beta$ and $\gamma$ are all scalar. (b): 
        The path diagram  representing mediation   with the density  as the mediator. The risk factor exposure $Z$ and psychological/physiological outcomes $Y$ are scalars, while the physical activity $M^{-1}(t)$ is a quantile function. Both the $\alpha(t)$ and the $\beta(t)$ pathways are represented by functions, while the $\gamma$ pathway is a scalar.}\label{fig:diag}
\end{figure}

Mediation analysis has been applied in many disciplines such as psychometric and behavioral sciences, and has received recent attention in statistics. 
Typically, a path diagram (Figure \ref{fig:diag} (a)) is used to illustrate the relationships among  three variables,  treatment $Z$,  outcome $Y$, and  mediator variable $M$. The causal path is often formulated by structural equation models (SEMs) with the coefficients representing causal effects.
Methods of mediation analysis  have been mostly focused on the case with a scalar mediator under the  linear structural equation
model \citep[e.g.,][]{ten2007causal,albert2008mediation,jo2008causal,sobel2008identification}. Recently,  researchers have extended the scope of mediation analysis to consider multiple mediators or more complex mediators such as images or functional data.  For example, \cite{preacher2008asymptotic} and \cite{vanderweele2014mediation} considered multiple mediators. \cite{sohn2019compositional} developed  a compositional mediation model where microbial composition is used as a mediator.  To deal with functional mediator, \cite{lindquist2012functional} extended SEMs to a linear functional structural equation model. 

This paper considers the setting when the possible mediator  is physical activity  that is  measured by modern wearable devices such as actigraphy and  is represented as minute-by-minute measures of activity levels.   As a motivating example,   the Individualized Comparative Effectiveness of Models Optimizing Patient Safety and Resident Education (iCOMPARE) trial is a national cluster-randomized trial that compared patient safety, education of trainees (both interns and residents), and intern sleep and alertness in internal medicine residency programs that were regulated by the 2011 duty-hour policies. In contrast, flexible policies had no limits on shift length or mandatory time off between inpatient shifts \citep{basner2019sleep}. 
During the 14-day iCOMPARE sleep study, interns wore an actigraph (model wGT3X-BT, ActiGraph) on the wrist of their non-dominant hand to track physical activities, and completed a brief survey on the smartphone each morning, including a sleep log, a score for sleep quality, the score on the Karolinska Sleepiness Scale (KSS), and a brief psychomotor vigilance test (PVT-B) \citep{basner2011validity}. One important question is to understand how flexible programs change interns' wake activity levels that affect sleep duration, sleepiness, and alertness of the interns.

To address this question, we propose a novel mediation model to study the indirect effect of flexible-duty policies transmitted through physical activity.
%Physical activity is recorded by wearable devices such as actigraphs, which collect minute-by-minute activity intensities continuously. 
To avoid the need of time registration and to reflect the daily activity level, activity trends are usually transformed into daily activity count distributions. Therefore, the mediator, physical activity, is in the form of a density. 
Due to the inherent constraints of densities, such as a constrained integral for density functions and monotonicity for quantile functions, commonly used mediation analysis methods with mediators in the  Euclidean space   are not directly applicable for such density  mediators.

To deal with the challenge of having densities as mediators, we use the Wasserstein metric to measure the distance between two distributions. Using  the isometry between the Wasserstein space and the space of quantile functions, we consider the path diagram in Figure \ref{fig:diag} (b) with the quantile function $M^{-1}(t)$ as the mediator and introduce a structural equation model for density . 
The treatment effect on distributions has been studied in \cite{lin2021causal} with some modeling assumptions on the  quantile functions. 
However, unlike the unrestricted functions, these assumptions may not be satisfied for the quantile functions due to  the monotonic increasing property.
To address this issue, we consider a general setting for $M^{-1}(t)$ and apply  the techniques introduced in \cite{han2019additive} to estimate the treatment effect on the density. To complete the SEM for density, we assume a functional linear model to model the effect of treatment and activity distribution on the outcome. 

The article is organized as follows. We introduce a mediation model for density  and discuss model assumptions and identifiability conditions in Section \ref{sec:mediation}.
Section \ref{sec:estandinfer} presents methods of estimating path coefficients and their covariance matrices. We also present a simple procedure for performing inference for the indirect effect and introduce a method for sensitivity analysis.
We examine our methods through simulations in Section \ref{sec:simulation} and an application to the iCOMPARE  trial data in Section \ref{sec:application}.

\ignore{
    \begin{figure}[!ht]
        \centering
        \includegraphics[width=0.75\textwidth]{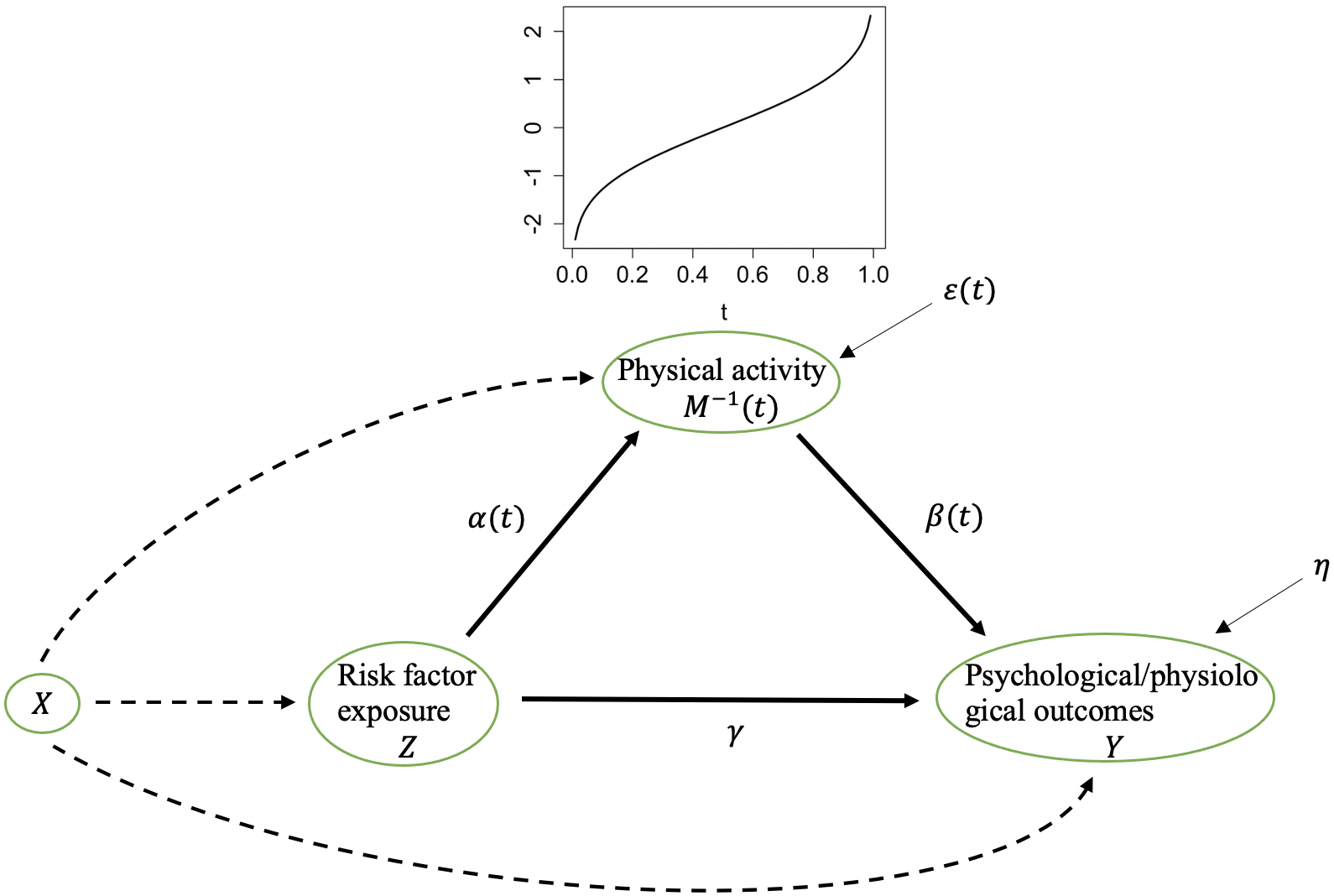}
        \caption{The path diagram used to represent the density  mediation framework. The risk factor exposure $Z$ and psychological/physiological outcomes $Y$ are scalars, while the physical activity $M^{-1}(t)$ is a quantile function. Both parameters $\alpha(t)$ and  $\beta(t)$ are represented by functions, while parameter  $\gamma$  is a scalar.}\label{fig:med}
    \end{figure}
    
}

\section{Mediation model and identification}\label{sec:mediation}

\subsection{Causal model with a density  as the mediator}\label{sec:c-model}
We consider the setting with a binary treatment or intervention. For subject $i~(i=1,\cdots,n)$, let $Z_i$ be the assignment variable with $Z_i=1$ if subject $i$ is assigned to treatment and 0 otherwise. Let $M^{-1}_i(t)$ be a quantile functional mediator at quantile $t$, $Y_i$ be a continuous  outcome, and $X_i\in R^d$ be a set of pre-treatment variables.
We adopt the potential outcomes framework to construct a causal model.
Denote by $M_i^{-1}(z,t)$ the potential quantile outcome under $Z_i=z$.
Denote by $Y_i(z,m^{-1}(t))$ the potential outcome under $Z_i=z$ and $M_i^{-1}(t)=m^{-1}(t)$.

Consider a given quantile $t$, under the stable unit treatment value assumption (SUTVA), the causal effects of $Z$ on $Y$ can be represented as the sum of two types of effects: (a) a mediated (or indirect) effect of $Z$ on $Y$ through $M^{-1}(t)$ and (b) an unmediated (or direct) effect of $Z$ on $Y$. For subject $i$, we have the following decomposition of the total treatment effect, 
\begin{align}\label{eq:yi}
    & Y_i(1,M^{-1}_i(1,t)) - Y_i(0,M^{-1}_i(0,t)) \notag \\
    = & \{Y_i(1,M^{-1}_i(0,t)) - Y_i(0,M^{-1}_i(0,t))\} + \{Y_i(1,M^{-1}_i(1,t)) - Y_i(1,M^{-1}_i(0,t))\} \notag \\
    = & \{Y_i(1,M^{-1}_i(1,t)) - Y_i(0,M^{-1}_i(1,t))\} +\{Y_i(0,M^{-1}_i(1,t)) - Y_i(0,M^{-1}_i(0,t))\}.
\end{align}
Since only one of the potential outcomes is observed in practice, the individual total treatment effect is unavailable and we seek to analyze average total treatment effect:
\[
\tau^{(c)} \equiv E\{Y(1,M^{-1}(1,t))-Y(0,M^{-1}(0,t))\}.
\]

We use the potential outcomes to express the causal relationship between $Z$, $M$, and $Y$ as: 
\begin{equation}\label{eq:m-c}
    M_i^{-1}(z,t) = \delta_1^{(c)}(t,x_i)+\alpha^{(c)}(t,x_i)z+\epsilon_i(z,t), 
\end{equation}
\begin{equation}\label{eq:y-c}
    Y_i(z,m^{-1}(t)) = \delta_2^{(c)} + \gamma^{(c)} z+\int\beta^{(c)}(t)m^{-1}(t)dt+x_i^T\xi+\eta_i(z,m^{-1}(t)),
\end{equation}
where $E(\epsilon(z,t)|X=x)=0$ and $E\{\eta(z,m^{-1}(t))|X=x\}=0$, $\delta_1^{(c)}(t,x_i)$ is the effect of covariates $x_i$ on the mediator at quantile $t$, $\alpha^{(c)}(t,x_i)$ is the covariate-adjusted treatment effect on the mediator at quantile $t$. In model \eqref{eq:y-c}, $\gamma^{(c)}$ measures the direct effect of the treatment, $\xi$ represents the effect of the covariates, and $\beta^{(c)}(t)m^{-1}(t)$ is a function that quantifies the effect of the density  mediator on the outcome. 
We assume a general setting for the mediator without any specific modeling assumptions on $\delta_1^{(c)}(t,x_i)$ and $\alpha^{(c)}(t,x_i)$.

Equation \eqref{eq:m-c} implies
$
E\{M^{-1}(1,t)-M^{-1}(0,t)|X=x\} = \alpha^{(c)}(t,x)
$
and the treatment effect on the mediator is
$
E\{M^{-1}(1,t)-M^{-1}(0,t)\} = E(\alpha^{(c)}(t,X)),
$
where the expectation is over $X$. 
The treatment effect $E(\alpha^{(c)}(t,X)$ is closely related to the distance between the two barycenters of the two densities of the mediators. Specifically, let $\mathcal{W}_2$ denote the 2-Wasserstein space and let $W_2$ denote the corresponding metric.
We define the means of potential mediators using Wasserstein barycenters:
\begin{align*}
    \mu_{at} =& \arg\min_{\nu\in\mathcal{W}_2} E(W_2^2(M(a,t),\nu)),\quad a=0,1,\\
    = & \arg\min_{\nu\in\mathcal{W}_2} E\|M^{-1}(a,t)-\nu^{-1}\|_2^2 = (E(M^{-1}(a,t)))^{-1}.
\end{align*}
We have $\mu_{at}^{-1}=E(M^{-1}(a,t))$ and
$
W_2(\mu_{0t},\mu_{1t}) = \|\mu_{0t}^{-1}-\mu_{1t}^{-1}\|_2=\|E(\alpha^{(c)}(t,X))\|_2,
$
which  shows that  the Wasserstein distance $W_2(\mu_{0t},\mu_{1t})$ is  the $L^2$ norm of $E(\alpha^{(c)}(t,X))$.

In model \eqref{eq:y-c}, we use the integration over all the quatiles $t\in[0,1]$ to measure the effect of the mediator on the outcome.
Under this model, we have 
\[
E\{Y(1,m^{-1}(t))-Y(0,m^{-1}(t))|X=x\} = \gamma^{(c)},
\]
\[
E\{Y(z,m^{-1}(t))-Y(z,m_*^{-1}(t))|X=x\} = \int\beta^{(c)}(t)(m^{-1}(t)-m_*^{-1}(t))dt.
\]
By averaging $\eqref{eq:yi}$ over all subjects,  the average total effect can be decomposed into the sum
of the average direct and indirect effect:
\begin{align*}
    \tau^{(c)} = & \gamma^{(c)} + E\{Y(1,M^{-1}(1,t))-Y(1,M^{-1}(0,t))\} \\
    = & \gamma^{(c)} + E\{Y(0,M^{-1}(1,t))-Y(0,M^{-1}(0,t))\}.
\end{align*}

\subsection{Structural equation model for density  as a mediator}
Since for any subject we cannot observe the two potential outcomes under both treatment and control simultaneously,  the treatment effects specified in Section \ref{sec:c-model} cannot be estimated from the observations. 
Therefore, we  propose the following SEM, which corresponds to the path diagram in Figure \ref{fig:diag} (b) and has estimable parameters:
\begin{equation}\label{eq:m-s}
    M_i^{-1}(Z_i,t) = \delta_1^{(s)}(t,x_i)+\alpha^{(s)}(t,x_i)Z_i+\epsilon_i(t), 
\end{equation}
\begin{equation}\label{eq:y-s}
    Y_i(Z_i,M_i^{-1}(Z_i,t)) = \delta_2^{(s)} + \gamma^{(s)}Z_i+\int\beta^{(s)}(t)M_i^{-1}(Z_i,t)dt+X_i^T\xi+\eta_i.
\end{equation}
The parameters of equations \eqref{eq:m-s} and \eqref{eq:y-s} are identified through the definitions $E(\epsilon(t)|Z=z,X=x)=0$ and $E(\eta|Z=z,M^{-1}(Z,t)=m^{-1}(t),X=x)=0$.

Using the SEM, we obtain
\begin{equation}\label{eq:alpha-s}
    E\{M^{-1}(1,t)|Z=1,X=x\} - E\{M^{-1}(0,t)|Z=0,X=x\} = \alpha^{(s)}(t,x),
\end{equation}
\begin{align*}
    &E\{Y(Z,M^{-1}(Z,t))|Z=1,M^{-1}(Z,t)=m^{-1}(t),X=x\}\\
    -&E\{Y(Z,M^{-1}(Z,t))|Z=0,M^{-1}(Z,t)=m^{-1}(t),X=x\} = \gamma^{(s)},
\end{align*} and 
\begin{align*}
    & E\{Y(Z,M^{-1}(Z,t))|Z=z,M^{-1}(Z,t)=m^{-1}(t),X=x\} \\
    -&E\{Y(Z,M^{-1}(Z,t))|Z=z,M^{-1}(Z,t)=m_*^{-1}(t),X=x\} = \int\beta^{(s)}(t)(m^{-1}(t)-m^{-1}_*(t))dt.
\end{align*}
Similarly, we define the total effect of $Z$ on $Y$ under the SEM by
\begin{align*}
    \tau^{(s)}\equiv & E\{Y(1,M^{-1}(1,t))|Z=1\} - E\{Y(0,M^{-1}(0,t))|Z=0\}. 
\end{align*}
Using \eqref{eq:y-s} and \eqref{eq:alpha-s}, we obtain
\begin{align*}
    \tau^{(s)}= & \gamma^{(s)}+\int\beta^{(s)}(t)[E(M^{-1}(1,t)|Z=1) - E(M^{-1}(0,t)|Z=0)]dt \\
    = & \gamma^{(s)}+\int\beta^{(s)}(t)E(\alpha^{(s)}(t,X))dt.
\end{align*}

Hence, under the SEM,  the total effect $\tau^{(s)}$ is decomposed  into two parts: the direct effect $\gamma^{(s)}$ and the indirect effect $\int\beta^{(s)}(t)E(\alpha^{(s)}(t,X))dt$.
Unlike the parameters of the causal model in Section \ref{sec:c-model}, the parameters of the SEM are identifiable from the observed data. 
In general, the causal and SEM parameters are not equal, implying that the latter should not be interpreted as causal effects.
We present assumptions needed next  under which the causal parameters are equal to the corresponding SEM parameters and thus are identifiable.

\subsection{Equivalence of  causal parameters and SEM parameters}
In order to make parameters under the causal model   equal to their counterparts under the SEM,   we need some assumptions. In a randomized study, it is reasonable to assume that treatment assignment is ignorable given pre-treatment variables.
\begin{assumption}\label{ass1}
    The treatment assignment is independent of the potential outcomes given pre-treatment variables:
    \[
    Y(0,M^{-1}(0,t)), Y(1,M^{-1}(1,t)), M^{-1}(0,t), M^{-1}(1,t) \ind Z|X.
    \]
\end{assumption}
Under Assumption \ref{ass1}, it is easy to show that $\tau^{(c)}=\tau^{(s)}$ and $\alpha^{(c)}(t,x)=\alpha^{(s)}(t,x)$.
However, in general, $\gamma^{(c)}\neq\gamma^{(s)}$ and $\beta^{(c)}(t)\neq\beta^{(s)}(t)$. 
To ensure the equality of the remaining parameters, a sufficient condition is that the mediator is ignorable with respect to the potential outcomes given pre-treatment variables.
\begin{assumption}\label{ass2}
    The mediator is independent of the potential outcomes given pre-treatment variables:
    \[
    Y(z,m^{-1}(t))\ind M^{-1}(z,t)|X.
    \]
\end{assumption}

Under Assumptions \ref{ass1} and \ref{ass2}, it is easy to show that $\gamma^{(c)}=\gamma^{(s)}$ and $\beta^{(c)}(t)=\beta^{(s)}(t)$, which is stated in the following theorem.

\begin{theorem}
    Under SUTVA and Assumptions \ref{ass1} and \ref{ass2}, the causal direct effect $\gamma^{(c)}$ and the causal indirect effect $\xi^{(c)}$ are identifiable and given by
    \begin{align*}
        \gamma^{(c)} = &\gamma^{(s)}, \\
        \xi^{(c)} = &\xi^{(s)} \equiv \int\beta^{(s)}(t)E(\alpha^{(s)}(t,X))dt.
    \end{align*}
\end{theorem}

\section{Parameter estimation and inference}\label{sec:estandinfer}
\subsection{Estimation of direct and mediating effects of densities on outcomes}\label{sec:estimation}
We first consider  estimating the treatment effect on the mediator, i.e., $E\{\alpha^{(s)}(t,X)\}$.
For a quantile function $M^{-1}(t)$, we denote by $f_M$ its density function. 
The inherent constraints of quantile functions make methods of functional data analysis in a Euclidean space inapplicable. 
To address this issue, following \cite{han2019additive}, we consider a transformation  to map a quantile function into an unrestricted function in the $L^2$ space, and   deploy an additive functional regression model to fit the responses in the unrestricted space, finally transforming back to density space. 
For instance, we use the log quantile density transformation:
\[
\psi:\mathcal{F}\rightarrow L^2,\quad \psi(M^{-1}(t))=-\log\{f_M(M^{-1}(t))\}
\]
for a collection of quantile functions $\mathcal{F}$.
Other transformations that satisfy certain criteria can  also be used (See \cite{petersen2016functional} for more discussions). 

After the transformation, we can  apply different  estimation techniques to transformed functions.  In this paper, 
we adopt an additive regression model introduced in \cite{han2019additive}.
Let $r$ denote a zero-mean error process  in the $L^2$ space, with i.i.d.\ realizations $r_i$ that are independent from all other random elements in the regression model
\[
\psi(M_i^{-1}(t)) = g_0(t) + \sum_{j=1}^dg_j(t,x_{ij}) + g_{d+1}(t,z_i) + r_i(t),
\]
where $g_0$ is an unknown univariate function satisfying $g_0(t)=E\{\psi(M^{-1}(t))\}$ and $g_j(t,x)~(j=1,\cdots,d+1)$ are unknown bivariate  functions satisfying $E(g_j(t,X))=0$. 
We have
\[
E\{\psi(M^{-1}(t))|Z,X\} = g_0(t)+\sum_{j=1}^dg_j(t,X_{j}) + g_{d+1}(t,Z).
\] 
We estimate the  functions $g_j~(j=0,\cdots,d+1)$  using the method that involves  an extended version of smooth backfitting   \citep{han2019additive}.

Given the estimates $\hat g_j$ of $g_j$, the estimator of $E\{M^{-1}(Z,t)|Z,X\}$ is
\[
\hat m^{-1}(t|Z,X)\equiv\psi^{-1}\left(\hat g_0(t)+\sum_{j=1}^d\hat g_j(t,X_{j}) + \hat g_{d+1}(t,Z)\right),
\]
and the estimator of 
\[
E(\alpha^{(s)}(t,X)) = E\{E(M^{-1}(1,t)|Z=1,X) - E(M^{-1}(0,t)|Z=0,X)\}
\]
is 
\[
\hat \alpha^{(s)}(t)\equiv  \frac{1}{n}\sum_{i=1}^n\left( \hat m^{-1}(t|1,x_{i})- \hat m^{-1}(t|0,x_{i})\right). 
\]
%\[
%\frac{1}{n}\sum_{i=1}^n\left\{\psi^{-1}\left(\hat g_0(t)+\sum_{j=1}^d\hat g_j(t,x_{ij}) + \hat g_{d+1}(t,1)\right)-\psi^{-1}\left(\hat g_0(t)+\sum_{j=1}^d\hat g_j(t,x_{ij}) + \hat g_{d+1}(t,0)\right)\right\}.
%\]
%The variance of $E(\alpha^{(s)}(t,X))$ can be estimated by applying bootstrap.

We next consider the estimation of the parameters in \eqref{eq:y-s},  which consists of a functional predictor and a scalar response.
We express $\beta^{(s)}(t)$ as a linear combination of $K$ known basis functions $\phi_{k}(t)$, i.e.,
\[
\beta^{(s)}(t) = \sum_{k=1}^Kb_k\phi_k(t)= b^T\Phi(t),
\]
where $b=(b_1,\cdots,b_K)^T$ and $\Phi(t)=(\phi_1(t),\cdots,\phi_K(t))^T$. 
Denote by $\varrho=(\delta_2^{(s)},\gamma^{(s)},\xi^T,b^T)^T$ the parameters to be estimated. 
We estimate $\varrho$ by minimizing the following least-squares criterion
\begin{equation}\label{eq:beta1}
    \min_{\varrho}\sum_{i=1}^n\left(Y_i-\delta_2^{(s)}-Z_i\gamma^{(s)}-X_i^T\xi-\Lambda_i^Tb\right)^2,
\end{equation}
where $\Lambda_i=\int\Phi(t)M_i^{-1}(Z_i,t)dt$. 
%and $R=(R_{jk})_{(d+K+2)\times (d+K+2)}$ with $R_{jk}=0$ if $j$ or $k\in\{1,\cdots,d+2\}$ and $\int\phi_j''(t)\phi_k''(t)dt$ otherwise.
Let 
\[
Y = \begin{pmatrix}
    Y_1 \\
    \vdots \\
    Y_n
\end{pmatrix} \text{ and } G = \begin{pmatrix}
    1 & Z_1 & X_i^T & \Lambda_1^T  \\
    \vdots & \vdots & \vdots & \vdots \\
    1 & Z_n & X_n^T & \Lambda_n^T 
\end{pmatrix},
\]
then the solution of \eqref{eq:beta1} is 
$
\hat\varrho = (G^TG)^{-1}G^TY.
$
%We could also impose some penalty to ensure a smooth estimator for $\beta(t)$, which is omitted here.
%The variance of $\hat\varrho$ can be estimated by applying bootstrap.

\subsection{Inference  using bootstrap}
After obtaining estimates of the model parameters, we perform  statistical  inference on the indirect mediation effect.
The null hypothesis of no total indirect effect is given by
\begin{equation}\label{eq:infer1}
    H_0: \int\beta^{(s)}(t)E(\alpha^{(s)}(t,X))dt = 0,
\end{equation}
and the null hypothesis of no indirect effect at a given quantile $t$ can be formulated as
\begin{equation}\label{eq:infer2}
    H_0: \beta^{(s)}(t)E(\alpha^{(s)}(t,X)) = 0.
\end{equation}
While the global null \eqref{eq:infer1} indicates no overall mediation effect of activity density, test of mediation effect at a given quantile potentially provides insight on whether certain level of activity mediates the effect of the treatment. 

The null hypothesis \eqref{eq:infer1} is to test whether the whole activity quantile mediates the effect of a flexible duty-hour policy on a psychological/physiological outcome. The total indirect effect may be insignificant if the indirect effect is negative over an interval of $t$ and positive over the rest. In this case, we could not reject the null hypothesis \eqref{eq:infer1} as the positive indirect effect part is offset by the negative part, while we may reject the null hypothesis \eqref{eq:infer2} at some values of $t$. 
Therefore, it is of practical value to investigate both the total indirect effect by \eqref{eq:infer1} and the local indirect effect at a given $t$ by \eqref{eq:infer2}. 

% We propose two methods: an extension of the Sobel test \citep{sobel1982asymptotic} and a bootstrap method.

% To extend the Sobel test, we need to estimate the variance of $\int\hat\beta^{(s)}(t)\hat \alpha^{(s)}(t)dt\equiv\int\hat\theta(t)dt$.
% %Let $\hat\theta(t)=\hat\beta^{(s)}(t)\hat E(\alpha^{(s)}(t,X))$.
% %We first approximate the variance of $\hat\theta(t)$ by applying the first order asymptotic method described in \citep{bollen1987total} and obtain
% %\[
% %var(\hat\theta(t))\approx E(\hat\beta^{(s)}(t))^2var(\hat E(\alpha^{(s)}(t,X))) + E(\hat E(\alpha^{(s)}(t,X))^2) var(\hat\beta^{(s)}(t)).
% %\]
% Note that
% \[
% E\left(\int\hat\theta(t)dt\right) = \int E(\hat\theta(t)) dt
% \] 
% and
% \[
% E\left(\int\hat\theta(t)dt\right)^2 = \int E(\hat\theta(u)\hat\theta(v))dudv = \int E(\hat\beta^{(s)}(u)\hat\beta^{(s)}(v)) E(\hat\alpha^{(s)}(u)\hat\alpha^{(s)}(v)) dudv.
% \]
% The variance of $\int\hat\theta(t)dt$ can be obtained easily.

To avoid distributional assumptions, we will apply a bootstrap method to approximate the $p$-value. 
We only show the procedure for \eqref{eq:infer1}, and  the procedure for \eqref{eq:infer2} is similar.
We resample observations and denote by $\xi_{b}$ an estimated total indirect effect derived from a resampled dataset.
Since the distribution of $\xi_b-\xi^{(c)}$ approximates that of $\xi^{(c)}$ when the null hypothesis \eqref{eq:infer1} is true,
a $p$-value can be approximated by $2\sum_{b=1}^BI(\xi_b-\xi^{(c)}\geq\xi^{(c)})/B$ when $\xi^{(c)}\geq 0$ and $2\sum_{b=1}^BI(\xi_b-\xi^{(c)}<\xi^{(c)})/B$ when $\xi^{(c)}< 0$, where $I(\cdot)$ is the indicator function and $B$ is the number of bootstrap samples. 
If there are repeated measures for some subjects, we sample the measures within the same subject together when generating a resampled dataset.

\subsection{Sensitivity analysis}\label{sec:sensitivity}
The equivalence of causal parameters and SEM parameters relies on both Assumptions \ref{ass1} and \ref{ass2}. However, Assumption \ref{ass2} may not be satisfied if there exists omitted variables related to both the observed values of the mediator $M_i$ and the potential outcomes $Y_i$. 
To assess the sensitivity of our analysis to the violation of Assumption \ref{ass2}, we consider the following sensitivity parameter:
\[
\rho(t) \equiv corr(\epsilon_i(t),\eta_i).
\]

Equations \eqref{eq:m-s} and \eqref{eq:y-s} imply
\begin{align}\label{eq:betarho1}
    &Y_i(Z_i,M_i^{-1}(Z_i,t)) \notag \\
    =& \delta_2^{(s)} + \gamma^{(s)}Z_i+\int\beta^{(s)}(t)\left(\delta_1^{(s)}(t,x_i)+\alpha^{(s)}(t,x_i)Z_i+\epsilon_i(t)\right)dt+X_i^Tg+\eta_i \notag\\
    = & \delta_2^{(s)} + \gamma^{(s)}Z_i+\int\beta^{(s)}(t)\left(\delta_1^{(s)}(t,x_i)+\alpha^{(s)}(t,x_i)Z_i\right)dt+X_i^Tg+U_i,
\end{align}
where $U_i=\int\beta^{(s)}(t)\epsilon_i(t)dt +\eta_i$.
Let $\sigma_1^2=var(\eta_i)$ and $\Sigma(u,v)=E(\epsilon_i(u)\epsilon_i(v))$.
We have
% \begin{align}\label{eq:betarho1}
%  \sigma_2^2=  var(U_i) = & var(\int\beta^{(s)}(t)\epsilon_i(t)dt) + var(\eta_i) + 2cov(\int\beta^{(s)}(t)\epsilon_i(t)dt,\eta_i) \notag\\ 
%    = & \int\beta^{(s)}(u)\beta^{(s)}(v)\Sigma(u,v)dudv + \sigma_1^2 + 2\sigma_1\int\beta^{(s)}(t)\Sigma(t,t)^{1/2}\rho(t)dt,
% \end{align}
% and
\begin{align}\label{eq:betarho2}
    cov(U_i,\epsilon_i(t)) = & cov(\int\beta^{(s)}(t)\epsilon_i(t)dt,\epsilon_i(t)) +cov(\eta_i,\epsilon_i(t)) \notag\\
    = & \int\beta^{(s)}(u)\Sigma(u,t)du + \sigma_1\Sigma(t,t)^{1/2}\rho(t).
\end{align}
We use $\beta_\rho^{(s)}(t)$ to denote $\beta^{(s)}(t)$ satisfying \eqref{eq:betarho1} and \eqref{eq:betarho2} for a given $\rho(t)$.
The detailed estimation procedure is provided in Appendix.
Under Assumption \ref{ass1}, for a given correlation $\rho(t)$, the total indirect effect is identified and given by
$
\int\beta_\rho^{(s)}(t)E(\alpha^{(s)}(t,X))dt.
$, which can be used to assess the sensitivity of the estimates. 

\section{Simulation study}\label{sec:simulation}
In this section, we evaluate the performance of the proposed methods through simulation studies. 
Data are generated for 300 subjects, of which 150 are randomly assigned to a treatment group $(Z = 1)$ and 150 to a control group $(Z = 0)$.
For each subject $i$, we first generate functions $g_i(t)$ in the $L^2$ functional space and then let $M_i^{-1}(t)=\psi^{-1}(g_i(t))$.
The shape of $g_i(t)$ is based on a common time-varying function $h_i(t)$, which is defined according to our real dataset presented in the next section:
\[
h_i(t) = -cos(2\pi t)/2+x_{i1}t^2-x_{i2}t+5,
\]
where $x_{i1}\stackrel{i.i.d.}{\sim}N(10,1)$ and $x_{i2}\stackrel{i.i.d.}{\sim}N(12,1)$.

\begin{figure}[!h]
    \centering 
    \includegraphics[width=1.0\textwidth,
    height=0.2\textheight]{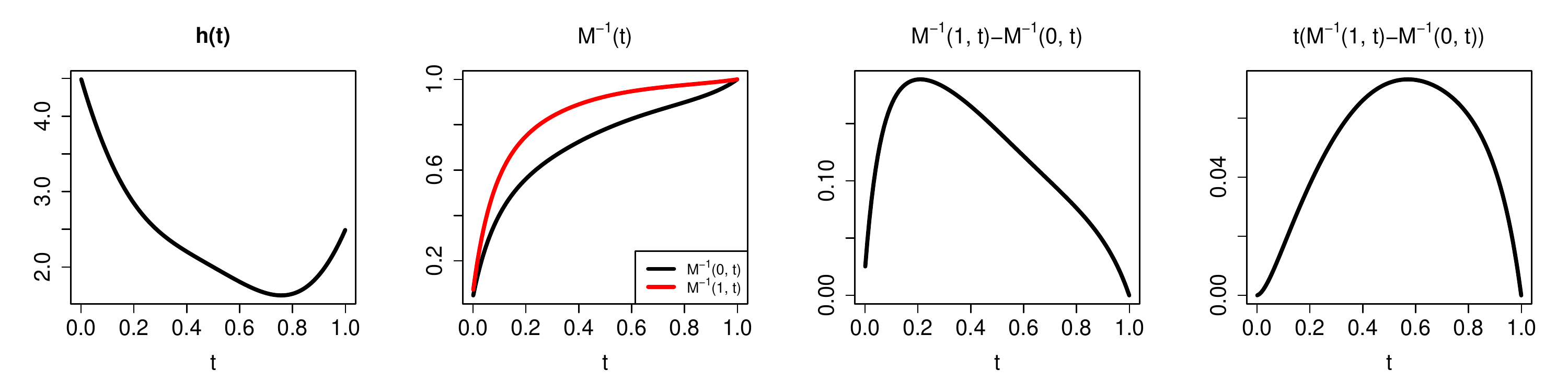}
    \caption{Functions used in simulations. From left to right: $h(t)=-cos(2\pi t)/2+10t^2-12t+5$; $M^{-1}(0,t)=\psi^{-1}(h(t))$, $M^{-1}(1,t)=\psi^{-1}(h(t)-2t)$; $M^{-1}(1,t)-M^{-1}(0,t)$; $t(M^{-1}(1,t)-M^{-1}(0,t))$.}\label{fig:oracle}
\end{figure}

Let $\epsilon_i(t)=\epsilon_{1i}\sin(\pi t)+\epsilon_{2i}\sin(2\pi t)$, where $\epsilon_{1i}\stackrel{i.i.d.}{\sim} N(0,1)$ and $\epsilon_{2i}\stackrel{i.i.d.}{\sim} N(0,1)$.
We consider the following four settings and and repeat the simulations  500 times.
\begin{itemize}
    \item Simulation 1: $M_i^{-1}(t)=\psi^{-1}(g_i(t))$, where $g_i(t)=\epsilon_i(t)$; 
    \\ $Y_i=0.05x_{i1}-0.05x_{i2}+Z_i+\eta_i$, where $\eta_{i}\stackrel{i.i.d.}{\sim} N(0,0.05^2)$.
    \item Simulation 2: $M_i^{-1}(t)=\psi^{-1}(g_i(t))$, where $g_i(t)= h_i(t)-2Z_it+\epsilon_i(t)$; 
    \\ $Y_i=0.05x_{i1}-0.05x_{i2}+Z_i+\eta_i$, where $\eta_{i}\stackrel{i.i.d.}{\sim} N(0,0.05^2)$.
    \item Simulation 3: $M_i^{-1}(t)=\psi^{-1}(g_i(t))$, where $g_i(t)= h_i(t)+\epsilon_i(t)$; 
    \\ $Y_i=0.05x_{i1}-0.05x_{i2}+Z_i+\int_0^1 t M_i^{-1}(t)dt+\eta_i$, where $\eta_{i}\stackrel{i.i.d.}{\sim} N(0,0.05^2)$.
    \item Simulation 4: $M_i^{-1}(t)=\psi^{-1}(g_i(t))$, where $g_i(t)= h_i(t)-2Z_it+\epsilon_i(t)$; 
    \\ $Y_i=0.05x_{i1}-0.05x_{i2}+Z_i+\int_0^1 t M_i^{-1}(t)dt+\eta_i$, where $\eta_{i}\stackrel{i.i.d.}{\sim} N(0,0.05^2)$.
\end{itemize}

In the first three settings, there is either no effect of $Z$ on $M$ or no effect of $M$ on $Y$, implying no indirect effect. We use Simulations 1--3 to examine the type 1 errors when no indirect effect is present.
To study the power  we design Simulation 4 with a significant indirect effect and present the shapes of related functions in Figure \ref{fig:oracle}.

\begin{figure}[H]
    \centering 
    \includegraphics[width=1\textwidth,
    height=0.20\textheight]{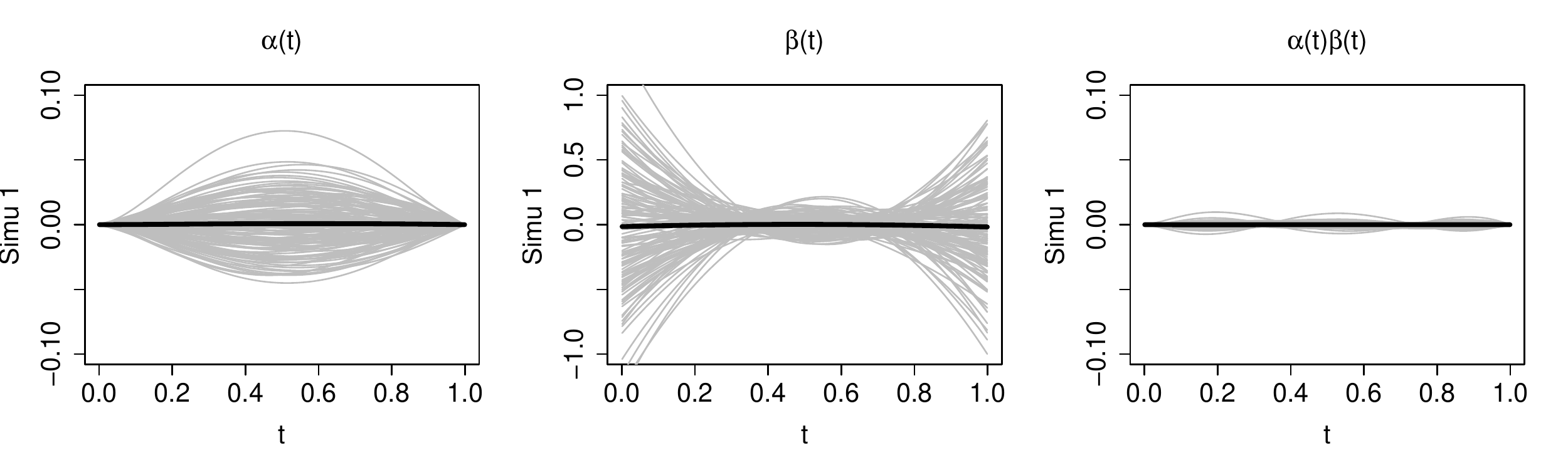}
    \includegraphics[width=1\textwidth,
    height=0.20\textheight]{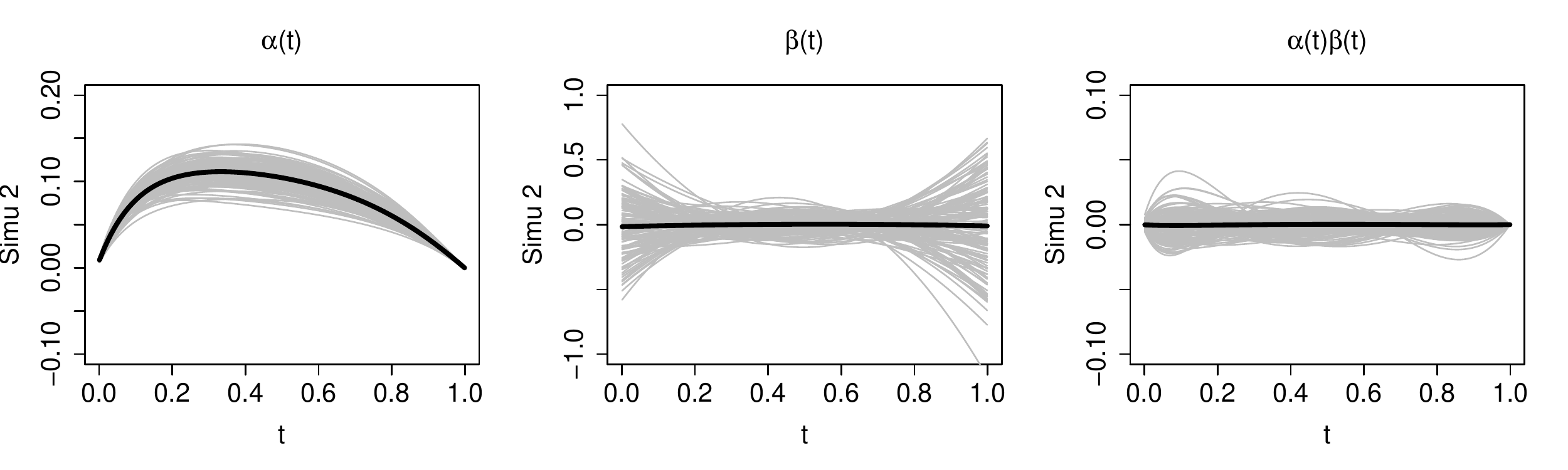}
    \includegraphics[width=1\textwidth,height=0.20\textheight]{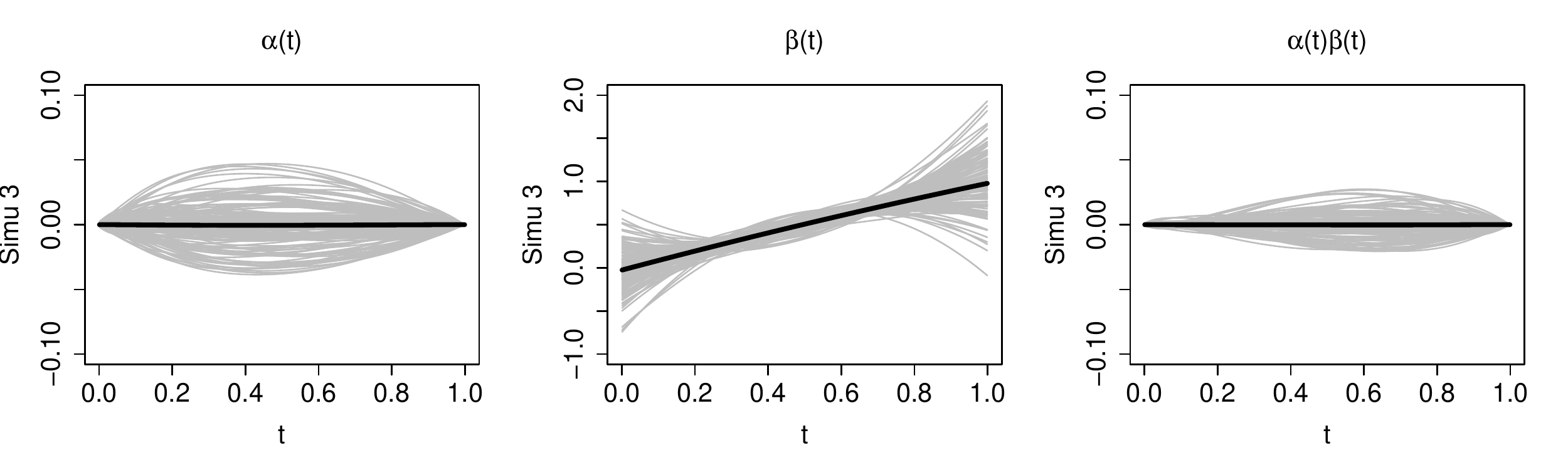}
    \includegraphics[width=1\textwidth,height=0.20\textheight]{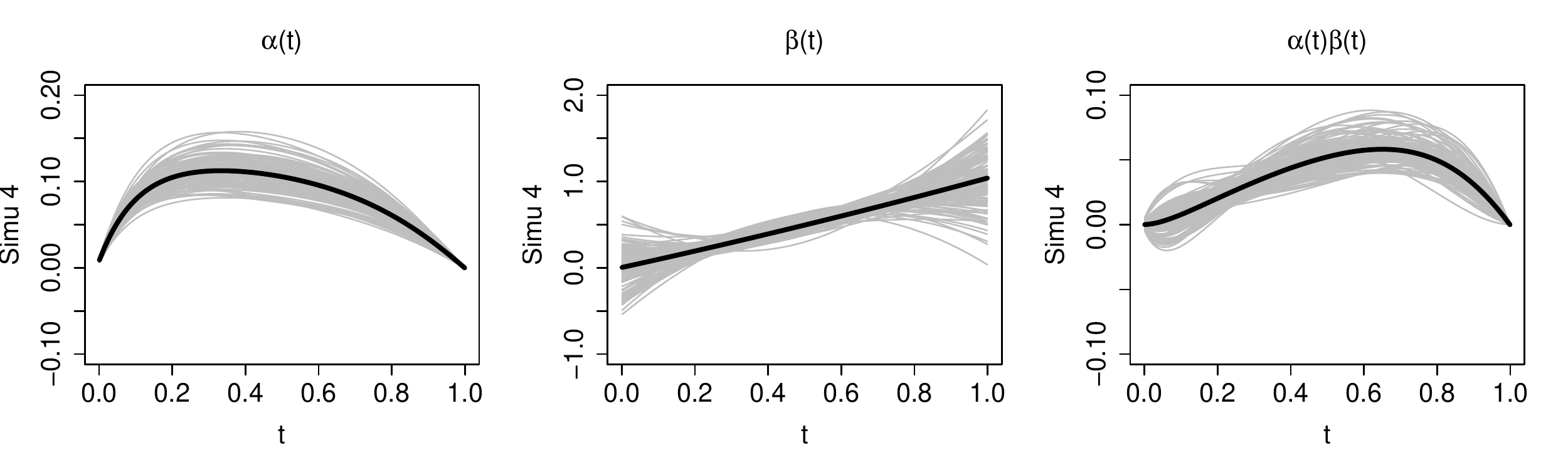}
    \caption{Estimates of $E\{\alpha^{(s)}(t,X)\},~\beta^{(s)}(t)$, and $\beta^{(s)}(t)E\{\alpha^{(s)}(t,X)\}$ in 500 runs under the four different models (top to bottom). The average functions are shown in bold.}\label{fig:simuest}
\end{figure}

The estimation results are shown in Figure \ref{fig:simuest}. We see under Simulation 1,  both estimates of $E(\alpha^{(s)}(t,X))$ and $\beta^{(s)}(t)$ are close to 0. 
Estimates of $E(\alpha^{(s)}(t,X))$ deviate  from 0 under Simulations 2 and 4 since there is an effect of $Z$ on $M$, and estimates of $\beta^{(s)}(t)$ are  different
from 0 under Simulations 3 and 4 due  to the effect of $M$ on $Y$. We observe  nonzero estimate of $\beta^{(s)}(t)E(\alpha^{(s)}(t,X))$ only under Simulation 4, which indicates the existence of indirect effect.

\begin{figure}[H]
    \centering 
    \includegraphics[width=1\textwidth,height=0.20\textheight]{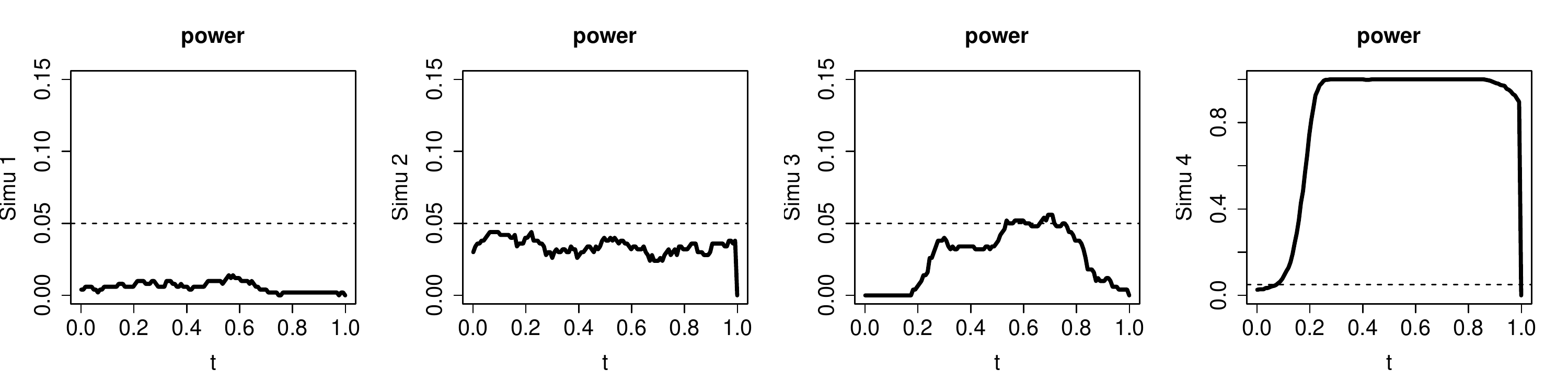}
    \caption{Type 1 errors  and empirical power of testing the mediation effects based on 500 runs under the four models, where models 1-3 are null model and model 4 has non-zero mediation effect.  }\label{fig:simupower}
\end{figure}

Figure \ref{fig:simupower} shows empirical power under the four simulation models. The results show  that the type 1 errors are  controlled for each of the quantile $t$ under Simulations 1--3 as we expected. The proposed test exhibits high power under Simulation 4, which is consistent with the estimate of $\beta^{(s)}(t)E(\alpha^{(s)}(t,X))$ shown in Figure \ref{fig:simuest}.

\section{Mediation analysis of iCOMPARE Trial}\label{sec:application}

\subsection{Data description and pre-processing}

We apply our method to the iCOMPARE clinical trial data \citep{basner2019sleep} mentioned in the Introduction to compare the flexible policies  (treatment)  and  standard policies  (control) in terms of their sleep deprivation, sleepiness and alertness.   iCOMPARE trial randomly assigned  398 interns to one of these two programs, of which 205 interns were randomly assigned to flexible programs  and others were assigned to standard programs. 
The study was approved by the Institutional Review Board of the University of Pennsylvania. Participants gave written informed consent prior to participation.
The interns in the treatment group worked under extended overnight shifts for most days and under regular day/night shifts for several days, while those in the control group worked under regular day/night shifts. 
During the 14-day iCOMPARE trial, every intern wore actigraphy to track physical activities, and completed a brief survey on the smartphone each morning, including a sleep log, a score for sleep quality (sleepqual), the score on the Karolinska Sleepiness Scale (KSS), and a brief psychomotor vigilance test (PVT-B). 
The 3-minute PVT-B is based on simple reaction time to stimuli that occur after randomly assigned 2-5 second inter-stimulus intervals \citep{basner2011validity}. It is a measure of behavioral alertness and has been shown to be very sensitive to sleep loss and circadian misalignment \citep{basner2011maximizing}.
The total number of sleep minutes (sleepmin) was derived based on data from actigraphy and sleep log information, and PVT response speed (pvtrespspeed; reciprocal response time) was recorded using   PVT-B.

One question of interest is to answer  whether a flexible duty-hour policies had an effect on the sleep related outcomes that was mediated through  shifting of physical activities. 
We  performed mediation analysis for each of the sleep related outcomes, including sleepmin, sleepqual, pvtrespspeed, and KSS. 
Specifically, the activity level in the last 24 hours (excluding off-wrist and sleep periods) was treated as the possible mediator. 
The minute-to-minute activity counts collected from actigraphies were summarized in a  vector. 
For subject $i$ on day $j$ ($j=1,\cdots,n_i$), let $A_{ij}=(a_{i1},\cdots,a_{il_{ij}})^T$ be the log-transformed activity counts, where $l_{ij}$ is the number of recorded activity counts for day $j$, and let $Y_{ij}$ be a sleep outcome measure corresponding to activity counts $A_{ij}$. 
We studied the daily activity by considering activity quantiles $M_{ij}^{-1}(t) = A_{ij}^{[l_{ij}t]}$, $t\in[0,1]$, where $A_{ij}^{[s]}$ denotes the $s$th order statistic of $A_{ij}$.
For each individual $i$, we summarized the activity data $M_{ij}^{-1}(t)$, $j=1,\cdots,n_i$, by obtaining the barycenter, denoted by $M_i^{-1}(t)$, and summarized the outcome data by taking the average of $Y_{ij}$, $j=1,\cdots,n_i$, denoted by $Y_i$. 
Our mediation analysis aimed to investigate how the flexible shift policies changed the average activity levels and average sleep related outcomes.

\begin{figure}[H]
    \centering 
    \begin{tabular}{c}
        %\vspace{-0.3in} 
        \includegraphics[width=\textwidth]{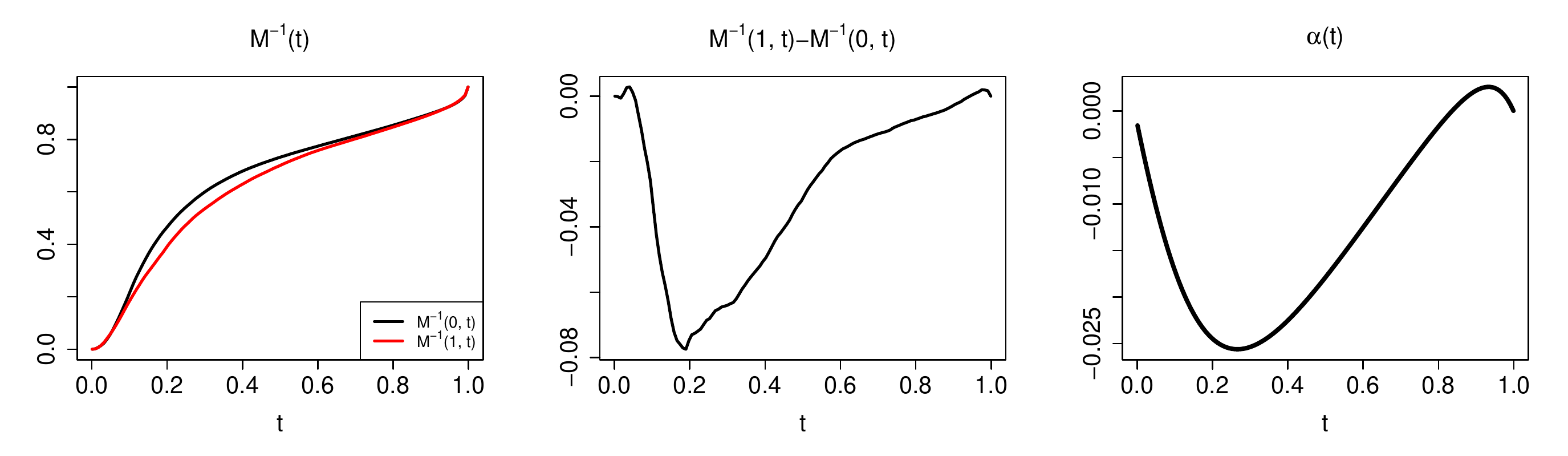}
        \end{tabular}
    \caption{Effects of flexible shift on activity level distributions.  The first column: barycenters of physical activity distributions for interns in the treatment and control groups; the second column: the difference between the two quantile functions; the third column: estimate of $E(\alpha^{(s)}(t,X))$ based on  the model \eqref{eq:m-s}.
    }\label{fig:real1}
\end{figure}

\subsection{Results of mediation analysis}

\begin{figure}[htbp]
\begin{center}
    \begin{tabular}{c}
        %\vspace{-0.3in}
        \includegraphics[width=\textwidth]{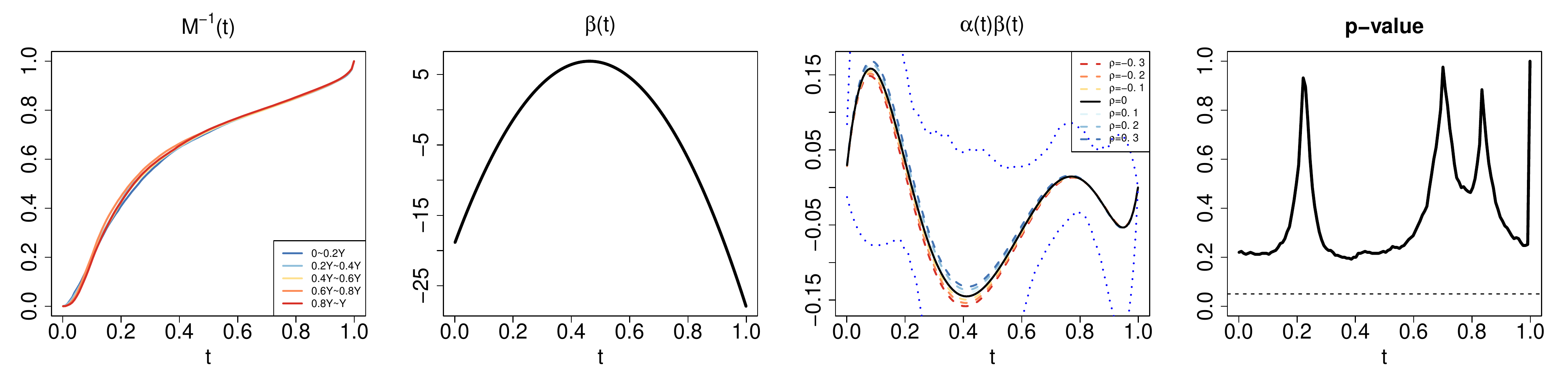}\\
        (a) Sleep min (sleepmin)\\
        %\vspace{-0.3in} 
        \includegraphics[width=\textwidth]{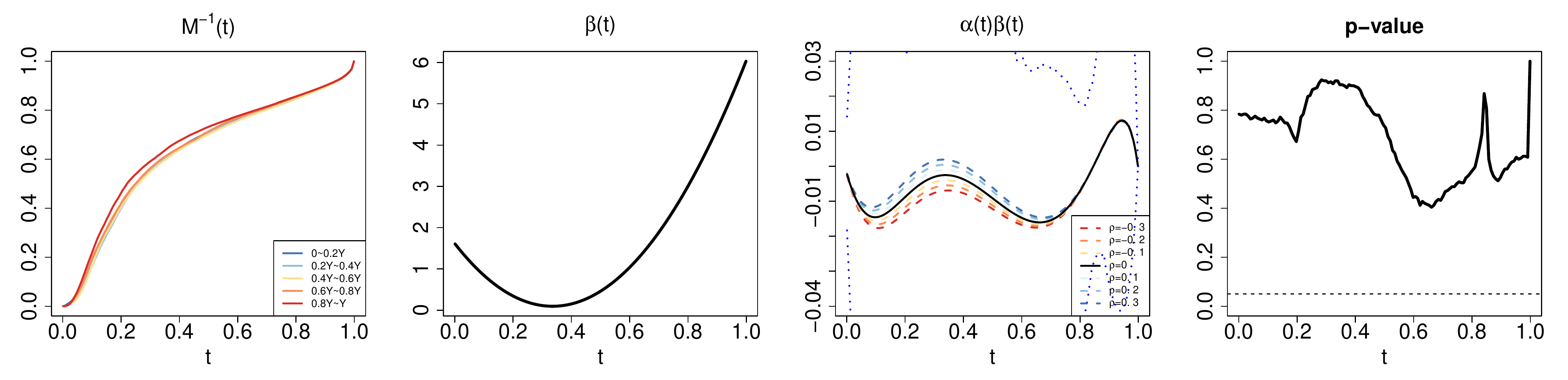}\\
        (b) Sleep quality (sleepqual)\\
        %\vspace{-0.3in}
        \includegraphics[width=\textwidth]{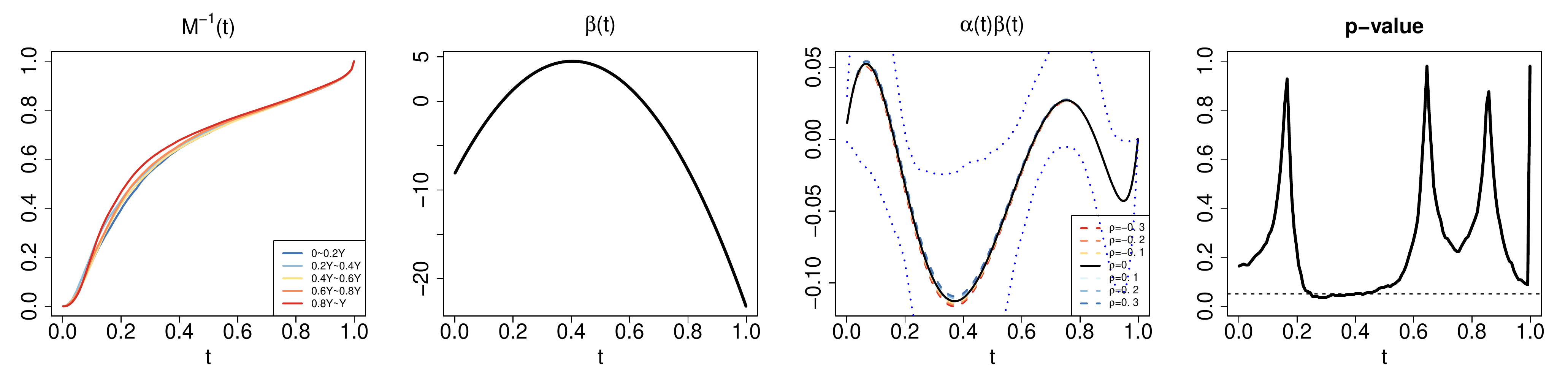}\\
        (c) PVT response speed (pvtrespspeed)\\
        %\vspace{-0.3in}
        \includegraphics[width=\textwidth]{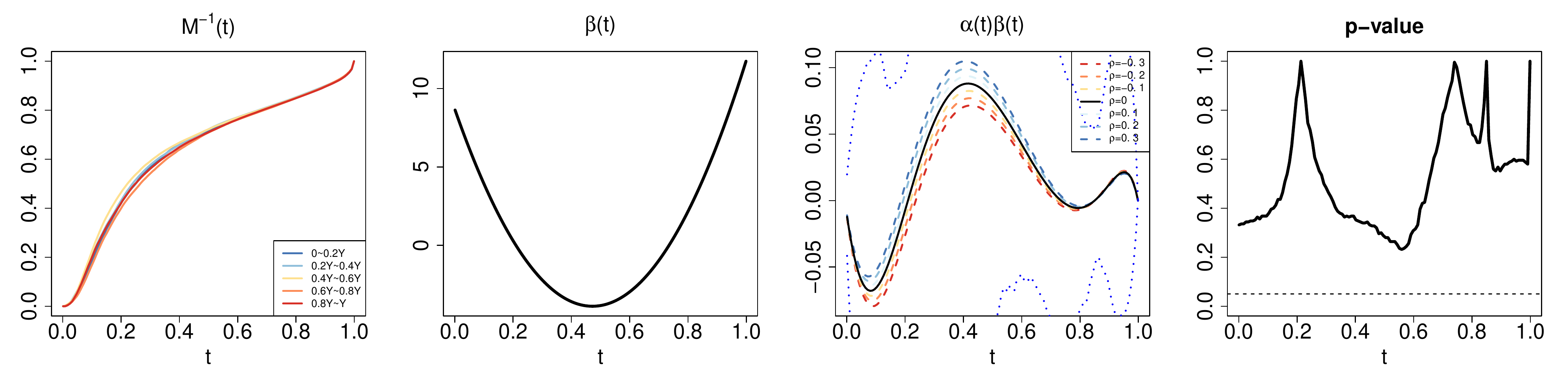}\\
        (d) Karolinska Sleepiness Scale (KSS)
    \end{tabular}
\end{center}
    \caption{Results of mediation analysis for four outcomes (from top to bottom): total number of sleep minutes (sleepmin), sleep quality (sleepqual), PVT response speed (pvtrespspeed), and Karolinska Sleepiness Scale (KSS), respectively.}\label{fig:real2}
\end{figure}

We  applied the path model shown in Figure \ref{fig:diag} (b), where the variable $Z_i=1$ if the subject $i$ was assigned to flexible policy program and 0 otherwise, the mediator $M_i^{-1}(t)$ was an activity quantile function, and the outcome $Y_{i}$ was one of the  sleep related outcomes.
To assess the effect of flexible shift policy on physical activity 
%based on the samples used to assess different sleep related outcomes,  
the first column of Figure \ref{fig:real1} shows the barycenters of physical activity distributions for interns in the treatment and control groups, showing some difference in mean distributions of the physical activity. The second column of Figure \ref{fig:real1} shows the difference between the two quantile functions, showing that the largest difference is at around $t=0.2$.  The third column shows the estimate of $E(\alpha^{(s)}(t,X))$ based on the model \eqref{eq:m-s}. We see the values are negative under almost the whole interval, indicating the negative effect of the flexible duty-hour policy on the physical activity, with the strongest effect at around $t=0.2-0.4$, i.e., low-to-moderate level of activity. 

The results of mediation analysis for the four outcomes, sleepmin, sleepqual, pvtrespspeed, and KSS, are shown in Figure \ref{fig:real2}. 
The plots in the first column show the barycenters of physical activity distributions under every 20 percent range of $Y$. We observe that interns with different values of pvtrespspeed have different activity distributions, while activity distributions under different values of sleepmin, sleepqual, and KSS seem to be similar, indicating the activity distribution is mainly associated with average PVT response speed. 
We present the estimates of $\beta^{(s)}(t)$ and $\beta^{(s)}(t)E\{\alpha^{(s)}(t,X)\}$ in the second and third columns for each of the four outcomes, respectively. 
The trends of $\beta^{(s)}(t)E(\alpha^{(s)}(t,X))$ for sleepmin and pvtrespspeed are very similar, indicating positive indirect effects under low activity regions and negative indirect effects under moderate-to-high activity regions.
We observe an opposite trend of $\beta^{(s)}(t)E(\alpha^{(s)}(t,X))$ for KSS with negative indirect effects under low activity regions and positive indirect effects under moderate-to-high activity regions.
The estimate of $\beta^{(s)}(t)E\{\alpha^{(s)}(t,X)\}$ for sleepqual is close to 0, indicating almost no mediation effect of activity distribution on sleep quality. 
The last column provides the approximate $p$-values from the bootstrap method for each of the quantile $t$. 
We observe significant indirect effects of the flexible duty-hour policy on pvtrespspeed through the physical activity under low-to-moderate activity regions. It shows no significant indirect effects on sleepmin, sleepqual, and KSS. After integrating the quantile $t$, we also observed a significant total mediation effect of activity distribution on pvtrespspeed ($p$-value $=0.05$).

\subsection{Sensitivity analysis}
In the real application, we can assume that random assignment in the Assumption \ref{ass1} holds, while the Assumption \ref{ass2} may not be satisfied. To assess the sensitivity, we estimate the indirect effects  assuming an error correlation of  $\rho(t)=-0.3,-0.2,\cdots,0.3$.
As shown in the third column in Figure \ref{fig:real2}, the indirect effects with nonzero correlations are within the 95\% bootstrap confidence intervals (blue dotted curved lines) of the estimated indirect effect with $\rho(t)=0$, indicating that our results are not sensitive to such correlations.

\subsection{Results from  repeated measurement analysis}
\begin{figure}[p]
    \begin{tabular}{c}
        %\vspace{-0.3in}
        \includegraphics[width=\textwidth]{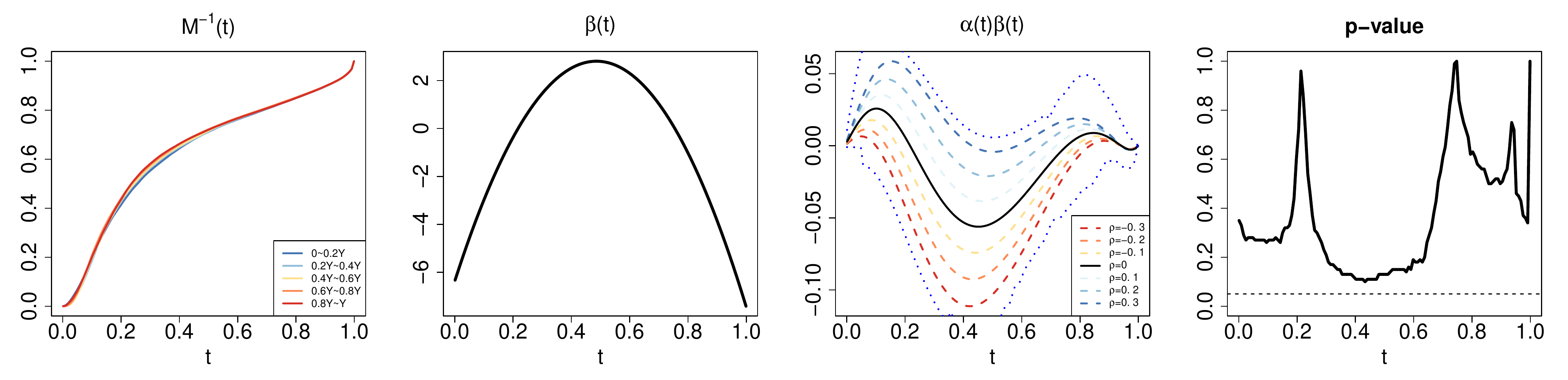}\\
        (a) Sleep min (sleepmin)\\
        %\vspace{-0.3in} 
        \includegraphics[width=\textwidth]{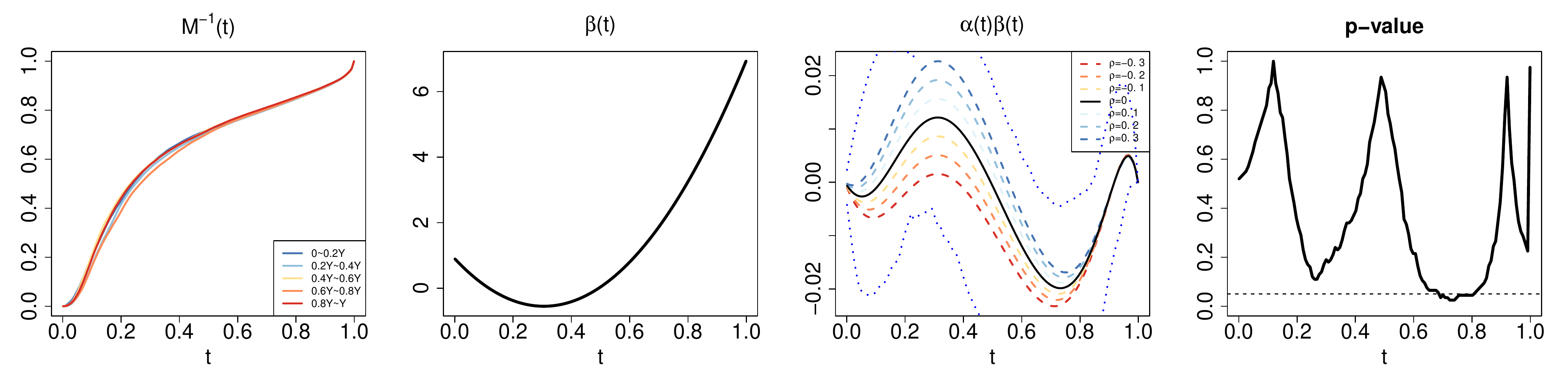}\\
        (b) Sleep quality (sleepqual)\\
        %\vspace{-0.3in}
        \includegraphics[width=\textwidth]{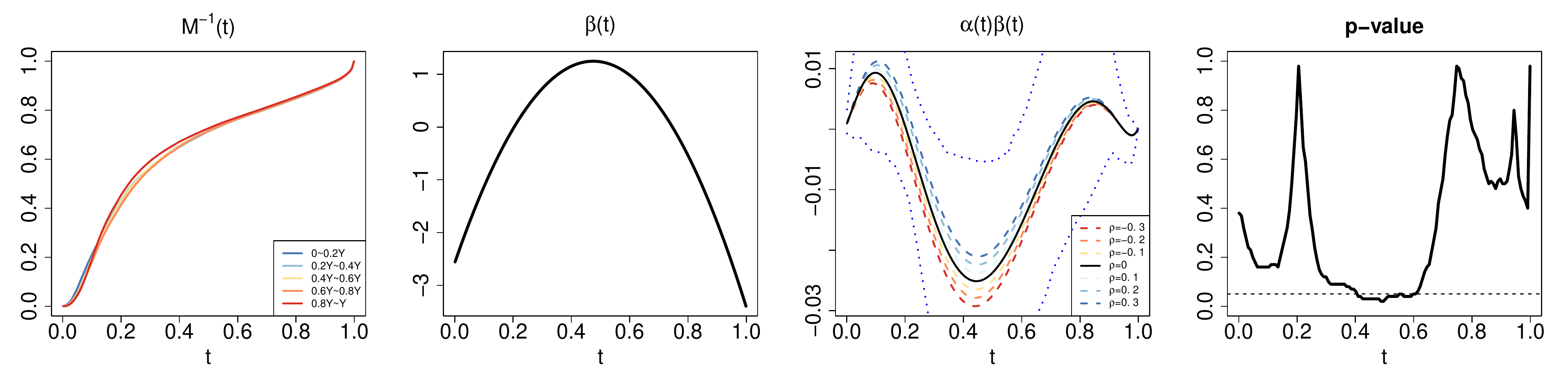}\\
        (c) PVT response speed (pvtrespspeed)\\
        %\vspace{-0.3in}
        \includegraphics[width=\textwidth]{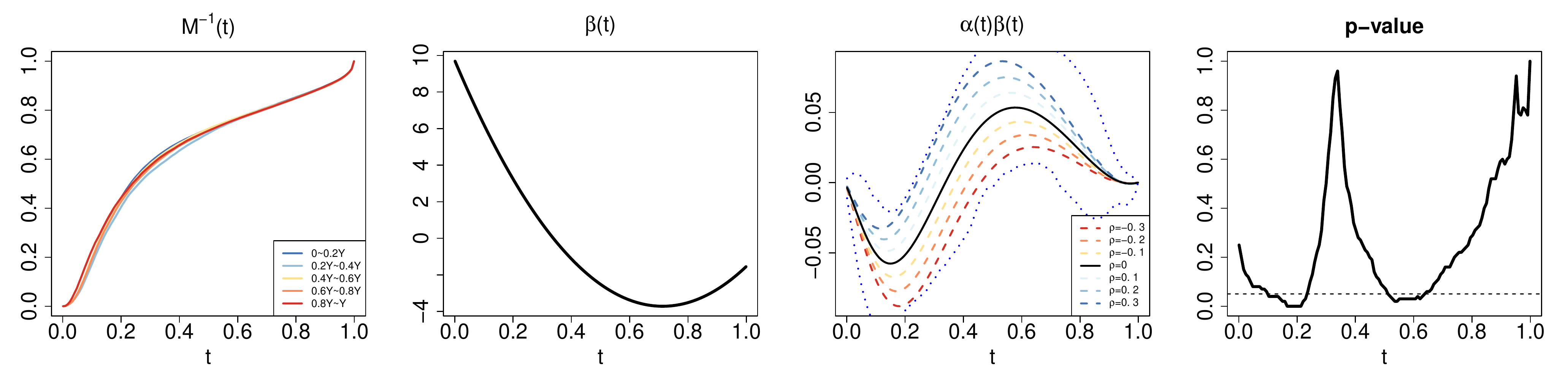}\\
        (d) Karolinska Sleepiness Scale (KSS)
    \end{tabular}
    \caption{Results of mediation analysis with repeated measures for four outcomes (from top to bottom): total number of sleep minutes (sleepmin), sleep quality (sleepqual), PVT response speed (pvtrespspeed), and Karolinska Sleepiness Scale (KSS), respectively.}\label{fig:realrep2}
\end{figure}

We also analyze the mediation effect in the iCOMPARE trial using data from all the repeated days during the trial. To evaluate the statistical significance, we take the within-subject repeated measures as a permutation unit in the bootstrap method.  The results are summarized in Figure  \ref{fig:realrep2}.
We observe a similar  estimate of $\beta^{(s)}(t)E(\alpha^{(s)}(t,X))$ using the repeated measures. However, 
the  confidence intervals in the third column of Figure \ref{fig:realrep2} are narrower,  which leads to more significant results as shown in the fourth column. 
The $p$-values in Figure \ref{fig:realrep2} indicate no significant indirect or mediation effects through activity on sleepmin.  
There are significant negative indirect effects of the flexible duty-hour policy on sleepqual and pvtrespspeed through the physical activity under high and moderate activity regions, respectively. 
For KSS, our analysis  shows significant negative indirect effects in the  low activity region and significant positive indirect effects in the moderate activity region.
After integrating the quantile $t$, we also observe a significant total mediation effect of activity distribution on pvtrespspeed ($p$-value $=0.03$).

\section{Conclusion and Discussion}

Motivated by increasing adaption of wearable device data in measuring individual's activities in population and clinical studies,  we have considered  the problem of mediation analysis where wearable device data are used as  a possible mediator to link treatment with an outcome.  In our formulation, we treat the minute-by-minute measures of activity as a density  that is used as a possible mediator to link random treatment or intervention  to the outcome. We have proposed a flexible model to link  the treatment  to  the density of the activity counts  and a model to link the density  of  the activity counts to the outcome. We have  developed the corresponding estimation and inference methods. 

In our analysis of the iCOMPARE trial,  since each intern had multiple days of data but had the same treatment assignment,  we summarized that activity data by obtaining the barycenter for each of the individuals and summarized the outcome data by taking the average of the observed outcomes.  
Our mediation analysis aimed to investigate how the flexible shift policies changed the distribution of the average activity levels and averages of  sleep related outcomes.  
In this work, we do not explicitly model the time-dependency of the activity and sleep outcomes and do not model how sleep outcomes of a given day affect the activity level of the next day. 
One interesting future research is to develop mediation analysis methods that  allow for such  time-dependent effects of the outcome on the mediators.

\section*{Acknowledgements}

The authors gratefully acknowledge supports from the NIH grants GM129781 and GM123056.
The iCompare trial was supported by grants U01HL125388 and U01HL126088 from the National Heart, Lung, and Blood Institute and grants from the American Council for Graduate Medical Education.

%  Here, we create the bibliographic entries manually, following the
%  journal style.  If you use this method or use natbib, PLEASE PAY
%  CAREFUL ATTENTION TO THE BIBLIOGRAPHIC STYLE IN A RECENT ISSUE OF
%  THE JOURNAL AND FOLLOW IT!  Failure to follow stylistic conventions
%  just lengthens the time spend copyediting your paper and hence its
%  position in the publication queue should it be accepted.

%  We greatly prefer that you incorporate the references for your
%  article into the body of the article as we have done here 
%  (you can use natbib or not as you choose) than use BiBTeX,
%  so that your article is self-contained in one file.
%  If you do use BiBTeX, please use the .bst file that comes with 
%  the distribution.  In this case, replace the thebibliography
%  environment below by 

\bibliographystyle{apalike}

\bibliography{distmed.bib}

%  If your paper refers to supporting web material, then you MUST
%  include this section!!  See Instructions for Authors at the journal
%  website http://www.biometrics.tibs.org

\appendix

%  To get the journal style of heading for an appendix, mimic the following.

\section{Procedure for sensitivity analysis in Section \ref{sec:sensitivity}}\label{app:sensitivity}
Given $\rho(t)$, we can estimate $\beta^{(s)}(t)$ as follows. \\
Step 1. Estimate $\delta_1^{(s)}(t,x_i)$, $\alpha^{(s)}(t,x_i)$, and $\epsilon_i(t)$ in \eqref{eq:m-s} by using the method in Section \ref{sec:estimation}. Then $\Sigma(u,v)=E(\epsilon_i(u)\epsilon_i(v))$ could be estimated. \\
Step 2. Given $\beta^{(s)}(t)$, let 
\[
\tilde Y_i(Z_i,M_i^{-1}(Z_i,t)) = Y_i(Z_i,M_i^{-1}(Z_i,t)) - \int\beta^{(s)}(t)\left(\delta_1^{(s)}(t,x_i)+\alpha^{(s)}(t,x_i)Z_i\right)dt.
\] 
Regress $\tilde Y_i(Z_i,M_i^{-1}(Z_i,t))$ on $(1,Z_i,X_i^T)$ and obtain the estimate of $U_i=\tilde Y_i(Z_i,M_i^{-1}(Z_i,t))-\delta_2^{(s)}-\gamma^{(s)}Z_i-X_i^Tg$. \\
Step 3. Obtain $\eta_i=U_i-\int\beta^{(s)}(t)\epsilon_i(t)$ and $\sigma_1^2=var(\eta_i)$. \\
Step 4. Estimate $\beta^{(s)}(t)$ by plugging $cov(U_i,\epsilon_i(t))$, $\Sigma(u,v)$, $\sigma_1$, and $\rho(t)$ into \eqref{eq:betarho2}. \\
Step 5. Repeat Steps 2--4 until convergence.
\end{document}